\documentclass[pra,showpacs,twocolumn,superscriptaddress,amsmath,amssymb]{revtex4-1}
 
 \pdfoutput=1

\usepackage{graphicx}
\usepackage{color}
\usepackage{dcolumn,amsmath}
\usepackage[normalem]{ulem}
\usepackage{xfrac}

\usepackage{dcolumn}% Align table columns on decimal point
\usepackage{bm}% bold math
\usepackage{amssymb}

\begin{document}

\title {Roaming pathways and survival probability in real-time collisional dynamics of cold and controlled bialkali molecules }

\author{Jacek K{\l}os}
\affiliation{Department of Physics, Temple University, Philadelphia, Pennsylvania 19122, USA}
\affiliation{Joint Quantum Institute, University of Maryland, College Park, Maryland 20742, USA}
\author{Qingze Guan} 
\affiliation{Department of Physics, Temple University, Philadelphia, Pennsylvania 19122, USA}
\author{Hui Li} 
\affiliation{Department of Physics, Temple University, Philadelphia, Pennsylvania 19122, USA}
\author{Ming Li} 
\affiliation{Department of Physics, Temple University, Philadelphia, Pennsylvania 19122, USA}
\author{Eite Tiesinga}
\affiliation{Joint Quantum Institute, University of Maryland, College Park, Maryland 20742, USA}
\affiliation{National Institute of Standards and Technology, Gaithersburg, Maryland 20899, USA}
\author{Svetlana Kotochigova}
\email{skotoch@temple.edu}
\affiliation{Department of Physics, Temple University, Philadelphia, Pennsylvania 19122, USA}

\begin{abstract}
Perfectly controlled molecules are at the forefront of the quest to explore chemical reactivity at ultra low temperatures. 
Here, we investigate for the first time the formation of the long-lived intermediates in the time-dependent scattering of cold bialkali 
$^{23}$Na$^{87}$Rb molecules with and without the presence of infrared trapping light. During the nearly 50 nanoseconds mean collision time of the 
intermediate complex, we observe  unconventional roaming  when for a few tens of picoseconds either NaRb or Na$_2$ and Rb$_2$
molecules with large relative separation are formed before returning to the four-atom complex.
We also determine the likelihood of molecular loss when the  trapping laser is present during the collision. 
We find that at a wavelength of 1064 nm the Na$_2$Rb$_2$ complex is quickly destroyed and thus that the
$^{23}$Na$^{87}$Rb molecules are rapidly lost. 
\end{abstract}

\maketitle

The availability of dilute ultracold gases of neutral  bialkali molecules in their absolute
ground state \cite{Ni2008, Liu2020, Takekoshi2014, Cornish2014, Park2015, Guo2016, Frauke2018, Ospelkaus2020}
and cold diatomic molecular ions \cite{Chen2011, Sullivan2012, Rellergert2013, Harter2013, Mohammadi2021}
is revolutionizing molecular science.  For heteronuclear dimers this stems from their
long-range and anisotropic dipole-dipole interactions.  Specifically, molecules with tunable
long-range interactions are an invaluable asset in the study of controlled chemical reactions
\cite{Krems2008, Martin2009, Dulieu2011, Balakrishnan2016, Hu2019, Hu2020}, while a gas of
such molecules can be used to explore exotic many-body physics \cite{Micheli2006, Yan2013}.
Moreover, their rotational states are excellent qubit states with natural entangling
interactions \cite{DeMille2002, Ni2018, Sawant2020, Hughes2020}.  These and other applications
based on either ultracold heteronuclear or homonuclear molecules require control of their
internal as well as external, motional degrees of freedom. This control is enabled by placing
the molecules in laser-generated optical potentials, {\it i.e.} optical traps, lattices, or tweezers, where they can 
be detected and manipulated.

From a chemistry point of view two classes of  ground-state alkali-metal  molecules
are relevant \cite{Zuchowski2010}. The first class are those molecules that are susceptible to
exothermic chemical reactions, such as all heteronuclear dimers that contain a Li atom and $^{40}$K$^{87}$Rb \cite{Zuchowski2010, Ospelkaus2010a,
Idziaszek2010, Kotochigova2010_1, DeMarco2019}. Total reaction rate coefficients for the
fermionic $^{40}$K$^{87}$Rb are on the order of $10^{-12}$ cm$^3$/s for the current
state-of-the-art nanokelvin temperatures when all molecules are prepared in the same spin
state. Rate coefficients are on the order of $10^{-10}$ cm$^3$/s  otherwise. Other bialkali
molecules are chemically stable with respect to molecule-molecule collisions. Examples are bosonic
$^{23}$Na$^{87}$Rb \cite{Guo2016, Ye2018},  $^{23}$Na$^{39}$K \cite{Ospelkaus2020}, and $^{87}$Rb$^{133}$Cs \cite{Takekoshi2014,
Cornish2014, Gregory2019}, and the fermionic $^{23}$Na$^{40}$K \cite{Park2015, Frauke2018}.
Their  lifetime in optical traps was expected to be only limited by the rate of collisions
with room-temperature molecules in the typical UHV vacuum environment used for these
experiments and expected to be several seconds. This would, for example, make possible the
simulation of novel many-body phases. Several recent experimental studies with these
``stable'' bosonic and fermionic molecules in optical traps, however,  observed a short
lifetime \cite{Takekoshi2014,Park2015,Ye2018,Gregory2019,Bause2021,Gersema2021} comparable to those of reactive molecules.

An explanation of this phenomenon, suggested by Ref.~\cite{Christianen2019}, relies on the
observation that electronically excited states of four-atom (tetramer) complexes become energetically
accessible to resonant absorption of photons from the trapping lasers. This leads to
uncontrolled spontaneous decay and loss of molecules from the trap. This loss
mechanism was confirmed by recent experimental observations  with $^{87}$Rb$^{133}$Cs
\cite{Gregory2020}. In addition, the rate of product formation in the reactive collision of
$^{40}$K$^{87}$Rb molecules was reduced by  trapping light \cite{Liu2020}.

The model of optical losses in collisions of fermionic NaK and KRb molecules in
Refs.~\cite{Christianen2019,Liu2020} was based on a comparison of the potential energies of
electronic ground and excited states of the four-atom complexes for a small set of geometries.
Laser excitation rate coefficients were obtained as a phase-space averaged sum over
excited states in time-independent perturbation theory.

The next logical step requires the time-dependent study of the quasi-classical collision dynamics of
alkali-metal molecules, where the electronic character of the few-body complex can
significantly change over only a few femtoseconds in synchrony with the motion of the nuclei.
Moreover, the intermediate complex can be long-lived, {\it i.e.} last significantly longer
then the characteristic rovibrational timescale of the reactants.  Finally, the dynamics is
typically barrierless \cite{Softley2009}.  As a result, the motion of the complex is expected
to be chaotic, thoroughly mixing all motional degrees of freedom, leading to a statistical
distribution of times between optical excitations and excitation probabilities, similar to the
statistical distributions of the reaction products observed in Ref.~\cite{Levine2009}.

In this paper, we investigate for the first time the real-time non-reactive collision of cold
millikelvin alkali-metal $^{23}$Na$^{87}$Rb molecules initialized in their absolute electronic
and ro-vibrational ground state in the presence of continuous-wave (cw) infrared 1064-nm optical radiation that
is off resonant for electronic excitations in the reactant dimers.  Our theoretical model
combines long-time quasi-classical trajectory (QCT) calculations using a global
six-dimensional ground-state Born-Oppenheimer potential energy surface (PES) of Na$_2$Rb$_2$
derived from an analytic dimer-in-molecule (DIM) model \cite{Ellison1963} with 
short-time QCT calculations using the ground  electronic PES derived from {\it on-the-fly}
density-functional-theory (DFT)  calculations.   The quantum nature of
the $v=0$ vibrational and $J=0$ rotational ground state of X$^1\Sigma^+$ NaRb molecules and
their cold external motion are accounted for by micro-canonical sampling over initial phase-space points
in position and momentum and computing the corresponding classical trajectories with Newton's
equations for the ground-state PES \cite{Hase1996}.  The relevant characteristics of the ground-state  
potential are described in Results and Methods.

The long-time QCT simulations with the easy-to-compute dimer-in-molecule PES have the primary
goal of determining the probability distribution for the duration or lifetime of the
intermediate tetramer complex but do not include laser radiation. These calculations are
discussed in Results and have led us to the realization that there exist two
types of trajectories.  Those with lifetimes less than 1 ns are quasi-one dimensional, while
those with lifetimes larger than 1 ns, corresponding to approximately 95\,\% of all
trajectories, undergo chaotic motion with mean collision time of $\approx50$ ns. The lifetimes
derived from these latter trajectories have Poisson statistics. 

Section Results describes our discovery of pervasive {\it roaming} trajectories in
$^{23}$Na$^{87}$Rb collisions that last more than 1 ns. Roaming trajectories or orbits
correspond to temporary large-amplitude motion of the colliding atoms in a weakly-bound
region of the tetramer potential. In our system roaming orbits correspond to the temporary
formation of well-separated homonuclear Na$_2$ and Rb$_2$ molecules and  
of NaRb+NaRb with or without  exchange of the two Na (or Rb) atoms.  
It is worth noting that, at our total energy, potential energies at conical intersections \cite{Domcke2011} between
the ground and first-excited Na$_2$Rb$_2$ states are  endothermic by about $hc\times 500$
cm$^{-1}$, where $h$ is Planck's constant and $c$ is the speed of light in vacuum. Conical intersections are also discussed in Results.

By convention, roaming states only correspond to large-amplitude atom configurations that are substantially different
from those obtained by  transition-state reaction pathway theories \cite{Bowman2014}. Over
the last twenty years, experimental and theoretical advances \cite{Townsend2004,
Christoffel2009, Takayanagi2011, Guo2013, Bowman2020} have made a convincing case for the
presence of roaming dynamics.  These studies, however, focussed on reactions with molecules
containing hydrogen and other light atoms and were restricted to collision temperatures above
one Kelvin \cite{Suits2008, Herath2011, Shepler2011, Bowman2014}.  Here, we have invigorated
the research on roaming by exploring collisions of heavy molecules at sub-Kelvin temperatures.

\begin{figure*}
	\includegraphics[scale=0.24,trim=0 0 0 0,clip]{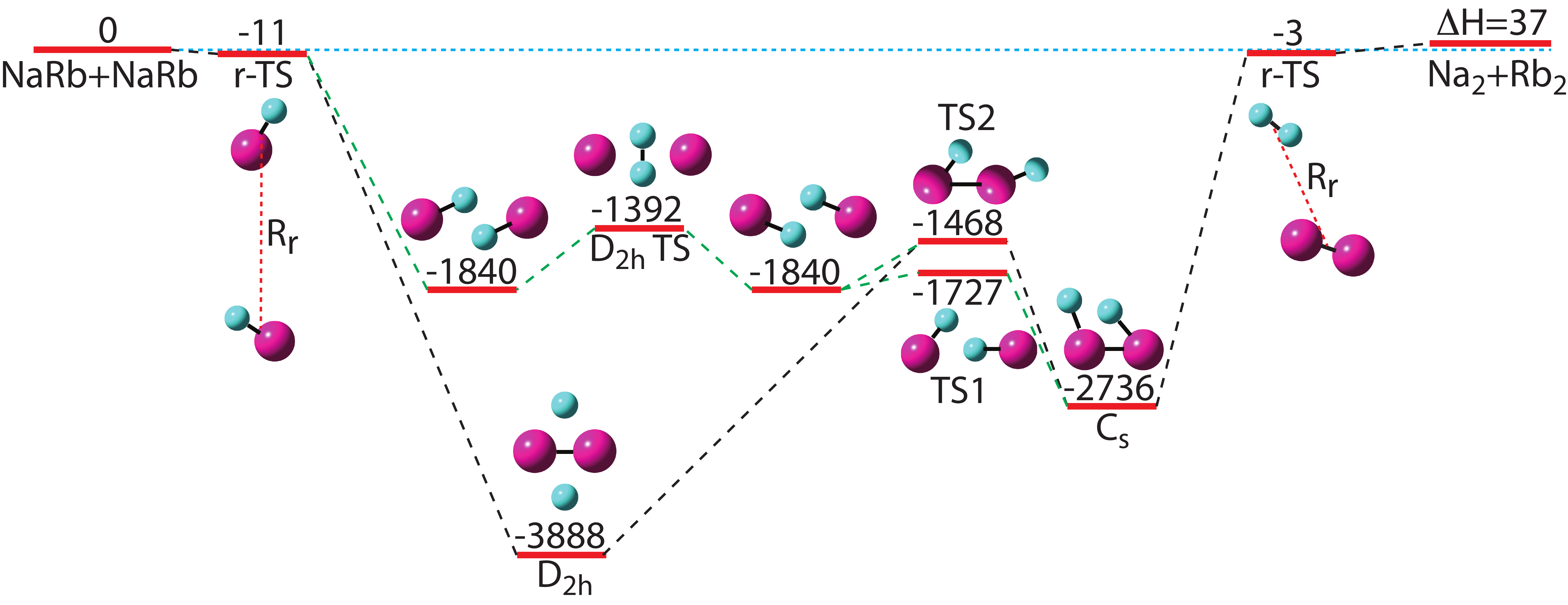}
\caption{Energies of stationary points  for the
$^{23}$Na$_2^{87}$Rb$_2$ ground-state potential based on coupled-cluster
calculations with all four atoms located in the same plane, {\it i.e.} have $C_s$ symmetry. All energies are given in units of $hc \times 1$ cm$^{-1}$.  The energy is zero for
two $^{23}$Na$^{87}$Rb molecules  at their equilibrium separation of the X$^1\Sigma^+$ state. 
This state is shown on the left-hand side of the figure.  The right-hand side shows the endothermic 
di-atomic $^{23}$Na$_2+^{87}$Rb$_2$ ground-state limit at the respective equilibrium separations. 
Other stationary states, minima and
saddle points, in the PES (red horizontal lines) are shown in between. Saddle points are labeled
by the abbreviations TS and r-TS corresponding to transition states and
roaming transition states, respectively.
For all stationary states a diagram shows the location of the four atoms.
 Na and Rb atoms are indicated by small cyan and large
magenta spheres, respectively.  
Dashed black and blue lines
connect stationary states and define  reaction paths.  Figure was prepared in Adobe Illustrator CS 6 with molecular structure pictures prepared using GaussView 5.09 for Mac.
	}
	\label{fig:diagram}
\end{figure*}

The short-time QCT simulations are based on {\it on-the-fly} DFT calculations of the ground-state
potential energy surface. Simultaneously, we use time-dependent density-functional-theory (TD-DFT) calculations to determine 
the splittings between excited- and ground-state potentials as well as  electric transition dipole moments to include the effects of the trapping light. These simulations follow Ref.~\cite{Hase2017} and are described in Results.  
Here,  atoms move classically with forces determined by the DFT ground electronic potential and can be excited when the energy difference of the ground and excited states  equals that of the energy of a photon in the trapping laser field.

Even when the DFT and TD-DFT potentials are only computed {\it on the fly} at the relevant classical positions of the 
atoms, the expense of electronic-structure
calculations is prohibitive and the computations are limited to evolution times of about 10
ps, much less than needed to determine the duration of a typical collision.  Even with this
short evolution time, however, it became evident that the 
ground-to-excited-state energy difference is often equal to that of the energy of an infrared photon, typically with wavelength of 1064 nm.  As a result, molecules moving
on the ground-state collisional complex can be promoted to excited states and then lost from
the trap.  We applied the Landau-Zener model \cite{Landau1932,Zener1932} at each resonant
geometry to determine an excitation probability.  Specifically, for a 1064-nm photon we find
that the mean time between optical excitations is about 0.2 ps  with a mean excitation
probability of $2\times 10^{-5}$ at a typical laser intensity of 10 kW/cm$^2$.
We have estimated the accuracy of the DFT and TD-DFT calculations  with additional  singles and doubles equation-of-motion coupled-cluster calculations (EOM-CCSD) at selected geometries along the classical paths.

The combination of the results from long- and short-time QCT simulations described in Results
also enabled us to determine the likelihood that $^{23}$Na$^{87}$Rb molecules are broken up  or destroyed by a laser during a collision.
We find that this is extremely likely for trapping lasers at a wavelength of 1064 nm.

\vspace*{0.5 cm}
\noindent
{\bf \large{Results}}\\
{\bf  Potential energy surface and reaction paths.} We  have performed  non-relativistic coupled-cluster (CC) calculations with single, double, and non-iterative triple
(CCSD(T)) excitations using the Karlsruhe def2-TZVPP basis set \cite{Weigend2005} in order to find stationary points and  reaction paths on the ground 
electronic state of  Na$_2$Rb$_2$. We correlated all inner and valence electrons in the CC calculations. This state has a zero total electron spin.  
The CC calculations show that the stationary points predominantly occur when the  four atoms are located at 
$C_s$ symmetries with all four atoms  in  a single plane.

Figure~\ref{fig:diagram} shows the potential energies at  stationary points with  $C_s$ symmetry as well as the
corresponding planar locations of the four atoms. 
The reactants NaRb+NaRb are shown on the left-hand side of the figure, while the
products Na$_2$+Rb$_2$ are shown on the right-hand side. The product energy  is only $hc\times37$ cm$^{-1}$ higher than that of the reactants and compares favorably with the 
spectroscopically determined value of $hc\times 47$ cm$^{-1}$  derived in Ref.~\cite{Ye2018}. 
For the energetics of this system it is also worth realizing that the quantum-mechanical zero-point energy of the individual
Na$_2$, NaRb, and Rb$_2$ dimers
are $hc\times 80$ cm$^{-1}$, $hc\times 53$ cm$^{-1}$, and $hc\times 29$ cm$^{-1}$
in their ground X$^1\Sigma_g^+$, X$^1\Sigma^+$, and X$^1\Sigma_g^+$ state, respectively.
This corresponds to a nearly identical combined zero-point energy of the reactants and products
that is significantly larger than the $hc\times37$ cm$^{-1}$ endothermicity of our CC calculations.
On the other hand, the difference between the combined zero-point energies is much smaller than the endothermicity. 

Figure \ref{fig:diagram} further shows four minima.
At the global minimum of the potential the atoms have  $D_{2h}$ symmetry as well. $D_{2h}$ symmetry corresponds to
planar geometries where the center of masses of the two Na atoms and  of the two Rb atoms coincide and the Na-Na and Rb-Rb interatomic axes make a $90^\circ$ angle. 
The extrema have multiple equivalent atom geometries or configurations with the same potential energy
reached by reflections through perpendicular planes or interchange of the Na atoms (or Rb atoms.)
Only for the  local minima at $hc\times-1840$ cm$^{-1}$ relative to the reactant state, do we show equivalent states as those 
will be helpful in describing reaction paths below.
Our  potential energies at the minima  agree well with the predictions in Ref.~\cite{Cote2012}.

The five saddle points or transition states in Fig.~\ref{fig:diagram} separate out into three ``conventional'' planar transition states (TS) and two roaming
transition states (r-TS). For both types of transition states  the curvature of the potential
along one or more normal-mode directions is negative, {\it i.e.}~has a so-called imaginary
frequency. Conventional or tight  transition states have imaginary frequencies that are on the
order of typical zero-point energies, here $\sim hc\times 100$ cm$^{-1}$ from the dimer
zero-point energies. Roaming transition states have at least one imaginary frequency that is
much smaller than these typical values.  Only one of the ``conventional'' transition states has
$D_{2h}$ symmetry.

The two r-TS in our system have potential energies close to the reactant and product states
and have geometries that closely resemble those of the reactant and product states.  They are
thus in the  threshold regions of the PES.  From Refs.~\cite{Harding2012, Guo2013} we learned
that such states can initiate dynamically distinct collisions.  For Na$_2$Rb$_2$, these transition states will
play an essential role in roaming collisional dynamics as discussed below.
The only out-of-plane stationary point (except for equivalent geometries reached by identical
atom interchange, reflections through the center of mass, etc.) is  a second-order saddle point at an energy of $-1070$ cm$^{-1}$
relative to the reactant state and has two directions along which the  curvature is negative. It is not
shown in Fig.~\ref{fig:diagram}.

Finally, Fig.~\ref{fig:diagram}  shows  planar reaction pathways. These reaction paths are one-dimensional
curves with $C_s$ symmetry connecting the reactants with the products along points of lowest
energy through intermediate saddle points and  local and global minima, {\it i.e.}  minimum
energy  paths (MEPs).  Two pathways connecting transition states and minima exist. The first
pathway passes through the global minimum with its $D_{2h}$ symmetry and crosses over the
tight TS2 transition state at $hc\times -1468$ cm$^{-1}$ to reach the  local minimum at
$hc\times-2736$ cm$^{-1}$ and the product state. The TS2 saddle point can also direct the first
pathway back to the  global minimum and reactant state.  The second pathway has a  flatter
energy landscape associated with the looser saddle points labelled $D_{2h}$TS and TS1 as well
as the two equivalent local minima at $hc\times-1840$ cm$^{-1}$.  Both pathways encounter our
two roaming transition states.

We end this subsection by noting that when not only the positions of the atoms lie in a 
single plane but also their momenta are directed in this plane then 
there exist no forces from the interatomic potentials that can break  $C_s$ symmetry. 
Similarly, if the positions and momenta of the four  atoms lie
on a single line, then they will always remain or move along this line.

\vspace*{0.5 cm}
\noindent
{\bf Collision time of the intermediate complex}. In this subsection we determine the mean duration or mean
collision time of the intermediate four-atom Na$_2$Rb$_2$ complex. 
We do so by computing an ensemble of classical trajectories of two colliding
$^{23}$Na$^{87}$Rb molecules using the Mercury/Venus96 code \cite{Hase1996, Garrido2019}. For these 
classical simulations we use an analytical ground-state PES based on the dimer-in-molecule (DIM) approach described in Methods. 
Essentially, this four-atom potential is constructed from the spectroscopically-accurate singlet X$^1\Sigma^+$ and triplet 
a$^3\Sigma^+$ pair-wise potentials for each of the dimers in the complex such that the 
total electron spin of the four doublet alkali-metal atoms is zero.
Coupled-cluster calculations, described in the previous subsection, are too computationally demanding within the Mercury/Venus96 code. 
The DIM potential can be quickly computed for any geometry of the four atoms, while still giving a reasonable description
of the four-body potential.

We set the total energy in the classical simulations equal to that of two  
$^{23}$Na$^{87}$Rb molecules, each in their $v=0,J=0$ ro-vibrational
state  plus a fixed relative kinetic energy of $k\times1$ mK,
where $k$ is the Boltzmann constant.
We use micro-canonical sampling for the initial relative orientation of the molecules and initial
relative momenta within each molecule given the constraints from  the total energy and zero point energies
of the $^{23}$Na$^{87}$Rb dimers.  We mainly present data for simulations with an impact parameter equal to
zero. At cold collision energies collisional lifetimes (and other ``global'' properties) are nearly independent of impact parameter 
up to  impact parameters  about five times larger than the equilibrium separation of ground state $^{23}$Na$^{87}$Rb. 
We postpone further discussion of the role of the impact parameter until the end of the last subsection of Results.  
Finally, the calculations are started at a molecule-molecule
separation of $R_s=57 a_0$, where the potential energy between the molecules is approximately
$hc\times -1$ cm$^{-1}$. The fixed propagation time step is 0.1 fs. Our calculations show that the total energy is conserved to 0.1 cm$^{-1}$. 

\begin{figure*}
	\includegraphics[scale=0.5,trim=0 20 100 20,clip]{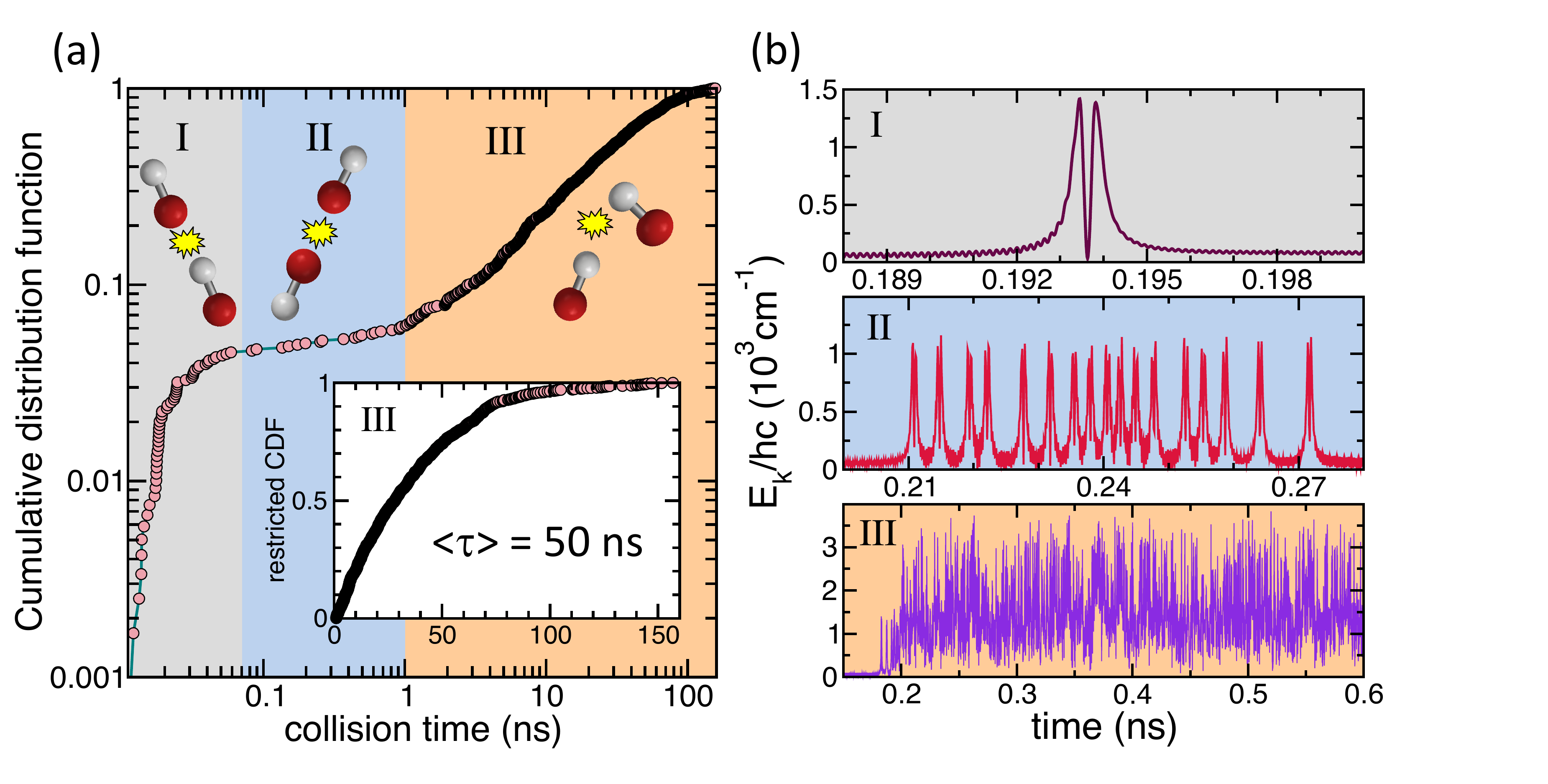}
	\caption{(a) Cumulative distribution function of the duration or collision time $\tau$ of the 
	collisional complex in cold $^{23}$Na$^{87}$Rb+$^{23}$Na$^{87}$Rb collisions determined
	from classical trajectory simulations starting with zero impact parameter. For each of the approximately 1200 trajectories the initial
	state is sampled assuming that the molecules are in the $v=0,J=0$ ro-vibrational ground state.
	Three classes of trajectories are  observed. They are indicated with colored bands and the roman numerals I, II, or III.
	The inset shows the restricted cumulative distribution function for trajectories versus $\tau$ in class III.
	(b) Total kinetic energies $E_{\rm k} $ as functions of time since the start of the simulation for one trajectory taken 
	from each of the three classes. Figures were prepared using Microsoft Power Point for Mac version 16.47 and Grace-5.1.22 software. 
	}
	\label{fig:lifetime}
\end{figure*}

For the purpose of defining the duration or lifetime of a collision, a collision starts when for the first
time the total kinetic energy is larger than two times the initial total kinetic energy. It ends
when for the last time the total kinetic energy is less than two times this initial total kinetic
energy. The time difference is the lifetime of a trajectory  $\tau$. The numerical simulations halt when the
separation between products reaches $R_f = R_s-0.5a_0$  (The choice of halting
separation slightly less than $R_s$ is an artifact of the fact that for cold collision and our
choice of $R_s$ products can not separate to infinity.  We have verified that the conclusions in this
section do not change when we use initial relative kinetic energies larger than $hc\times 1$
cm$^{-1}$.) We then analyze the geometry of the four-atom system. Mainly, the products are simply
those of the initial state or those where the two sodium atoms (or the two rubidium atoms) have
interchanged.  In rare cases the classical simulations lead to the homonuclear Na$_2$+Rb$_2$
products.   The endothermicity of the system, shown in Fig~\ref{fig:diagram}, is smaller than the
zero point energy of both NaRb+NaRb and Na$_2$+Rb$_2$. Thus with micro-canonical sampling the product
Na$_2$+Rb$_2$ can sometimes appear. In any quantum mechanical simulation this product cannot form.
They do not significantly affect our value for the collision lifetime.

Figure~\ref{fig:lifetime}(a) shows the cumulative distribution function (CDF) of lifetimes $\tau$
as determined from approximately 1200 zero-impact-parameter trajectories on a double logarithmic
scale.  The cumulative distribution function at argument $\tau$ is the probability that the
collision time is less than or equal to $\tau$.  We observe that there exist three distinct classes
of trajectories. Class I includes trajectories with $\tau<0.07$ ns. This group corresponds to
approximately 4\% of the trajectories. Class II includes the trajectories with $\tau$ between 0.07
ns and 1 ns and has fewer members. Class III consists of long-lived trajectories  with $\tau$ larger
than 1 ns with a majority of cases around 50 ns. 

The inset of Fig.~\ref{fig:lifetime}(a) depicts the cumulative distribution function for paths with
$\tau> 1$ ns. This restricted CDF is well represented by the function $1-\exp(-\tau/\langle\tau\rangle)$
corresponding to an exponential or Poissonian probability function of lifetimes.  In fact, the  mean
lifetime $\langle\tau\rangle$ is  50~ns from a least-squares fit.

Figure \ref{fig:lifetime}(b) shows  examples of the total kinetic energy as functions of time for one trajectory 
out of each of the three groups. The behavior of the kinetic energy for short-lived, class-I trajectories  is intriguing. 
An in-depth investigation showed that  for most of these trajectories
the two NaRb molecules approach each other in near collinear orientation and collide {\it head-to-tail}, as sketched in Fig.~\ref{fig:lifetime}(a), and their motion is  quasi-one-dimensional with small excursions along perpendicular directions weakly breaking collinear symmetry. 
Moreover, the atoms come
close only once, corresponding to the near zero kinetic energy around time 0.1936 ns. The existence of these
trajectories follows from the observation that when the positions and velocities of the atoms are in a single
common direction the forces that would move the atoms out of alignment are zero.
The relatively-large 4\% probability of such collisions is due to the small positive curvature
of the potentials along all transverse directions. The smaller this curvature or frequency, the
larger the acceptance angle for  quasi-1D collisions.
The rare class-II trajectories are initially head-to-head collisions with larger rotational
energy leading to their multiple re-collisions with various orientations. 

Finally, in long-lived, class III collisions  the atoms undergo chaotic  motion  with all orientations, geometries, and velocities 
appearing at random times. Short intervals of regular motion, however, do appear once in a while and will be discussed in the next 
section.

\vspace*{0.5 cm}
\noindent
{\bf Roaming dynamics and conical intersections.} Analysis of the long-lived, class III trajectories discussed in the previous section has led us to the observation of unconventional  {\it roaming} pathways, previously observed in unimolecular and atom-molecule reactions with light atoms \cite{Christoffel2009, Bowman2011, Guo2013}. A main characteristic of  roaming pathways is that they  go through  near-threshold roaming transition states (r-TS). They  exhibit  bond stretching with  large amplitude  relative motion of molecules.
\begin{figure} 
	\hspace*{-2mm}\includegraphics[scale=0.41, trim=0 0 0 0, clip]{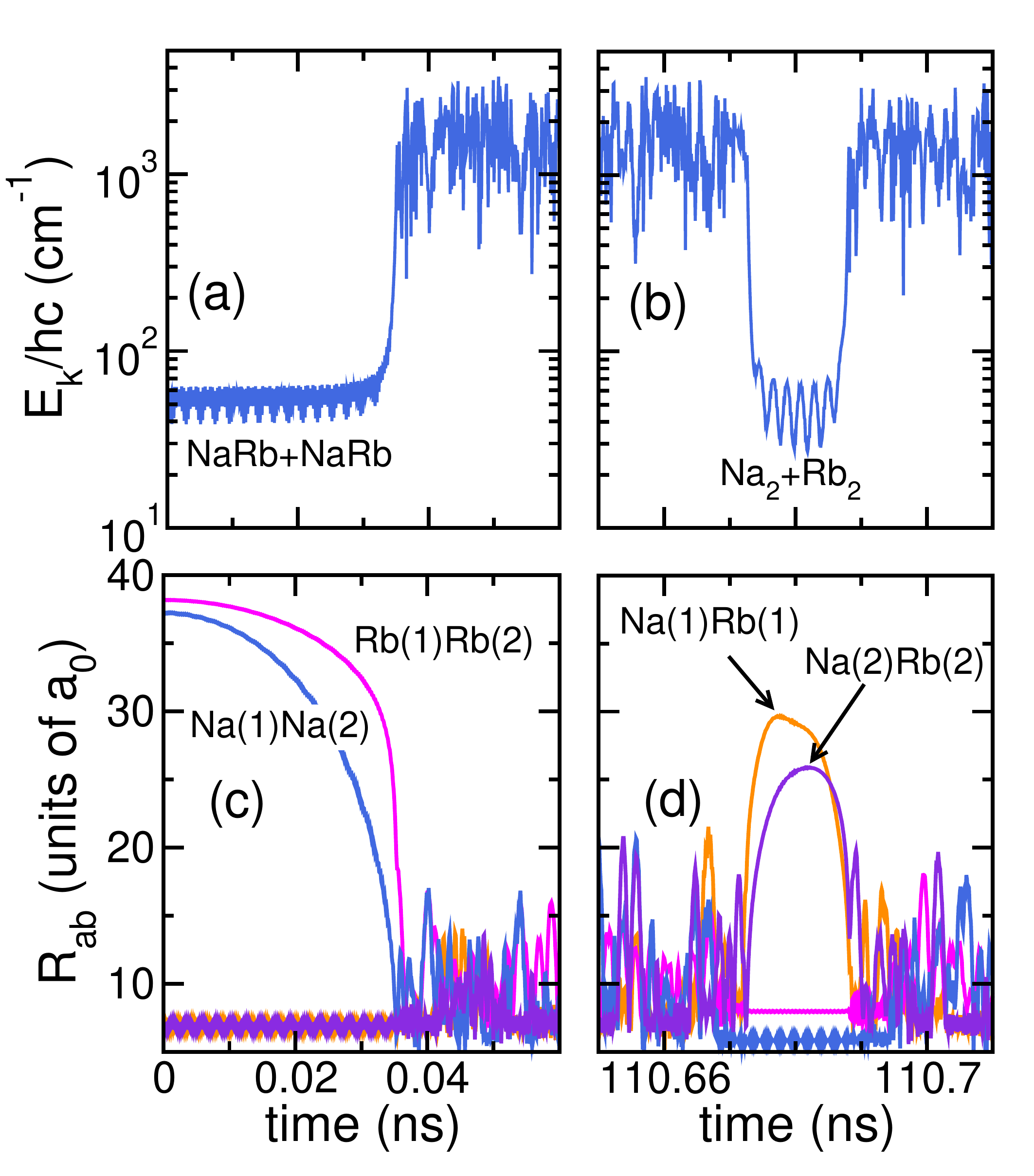} 
	\caption{Total kinetic energy, $E_k$, and interatomic separations, $R_{ab}$, as functions of time for a zero-impact-parameter trajectory with a lifetime 
	of 200 ns. 
Panel (a) shows $E_k$ for the first 0.06 ns of the association of two NaRb into a tetramer complex. Panel (b) shows a roaming 
event that occurs  110.66 ns into the collision. Panels (c) and (d) show the  relevant pair separations between the Na and 
Rb atoms as functions of time for the time intervals in panels (a) and (b), respectively.  Only pair separations with the largest values 
have been labeled. The line colors in the two panels
have the same meaning. Here, roaming leads to the temporary formation of homonuclear Na$_2$ and Rb$_2$. Figures were prepared using Grace-5.1.22 software.}
	\label{fig:roam}
\end{figure}

We observe multiple appearances of roaming pathways in the complex-forming Na$_2$Rb$_2$  trajectories. Motion in the 
complex is once in a while interrupted by dimer molecules moving away from the complex with large amplitude 
angular motion. This lasts for a few tens of ps  after which they return to again form a four-atom intermediate. This process  
continues until  products or reactants have enough kinetic energy to fly away from each other. An example of such  
event is shown in Fig.~\ref{fig:roam}. Panels (a) and (b) of this figure show the total kinetic energy, $E_k$, of a long-lived 
trajectory with lifetime $\tau=200$ ns  as functions of time. Panel (a) shows a time slice just at the beginning of the  association of 
two NaRb into a tetramer. Panel (b) displays a roaming event  in the trajectory. Panels (c) and (d) 
show the relevant pair separations. 
The bond between Na and Rb atoms is temporarily weakened, their separations are stretched out to $\approx 30 a_0$, while at the 
same time Na$_2$ and Rb$_2$ molecules are formed near the roaming-TS on the right-hand side of Fig.~\ref{fig:diagram}. 
These homonuclear dimers  slowly roam about each other in the 
van der Waals region of the potential. After about seven vibrational periods of the homonuclear dimers 
the roaming molecules return back to reform the tetramer.
In other trajectories with long lifetimes, we have also observed roaming to NaRb+NaRb over similarly brief time periods
corresponding to the  roaming-TS on the left-hand side of Fig.~\ref{fig:diagram}.

Previous studies of roaming with hydrogen-containing molecules \cite{Nakamura2014, Maeda2015} emphasized the  
role that seams of conical intersections (CIs) between two PESs \cite{Domcke2011}  play. Here, CIs are the geometries where the potential energies of molecular states with the same properties under symmetry operations are equal. The existence of CIs in  alkali-metal tetramers was previously predicted  by 
Refs.~\cite{Wallis2009, Zuchowski2010}.  For a better understanding of roaming in Na$_2$Rb$_2$, we studied the potentials of the 
singlet ground state and the first singlet excited state  within the dimer-in-atom model (See Methods for the calculation 
of the excited-state potential.) The dimer-in-atom model puts all CIs at $D_{2h}$ geometries. 

\begin{figure}
      \includegraphics[scale=0.38]{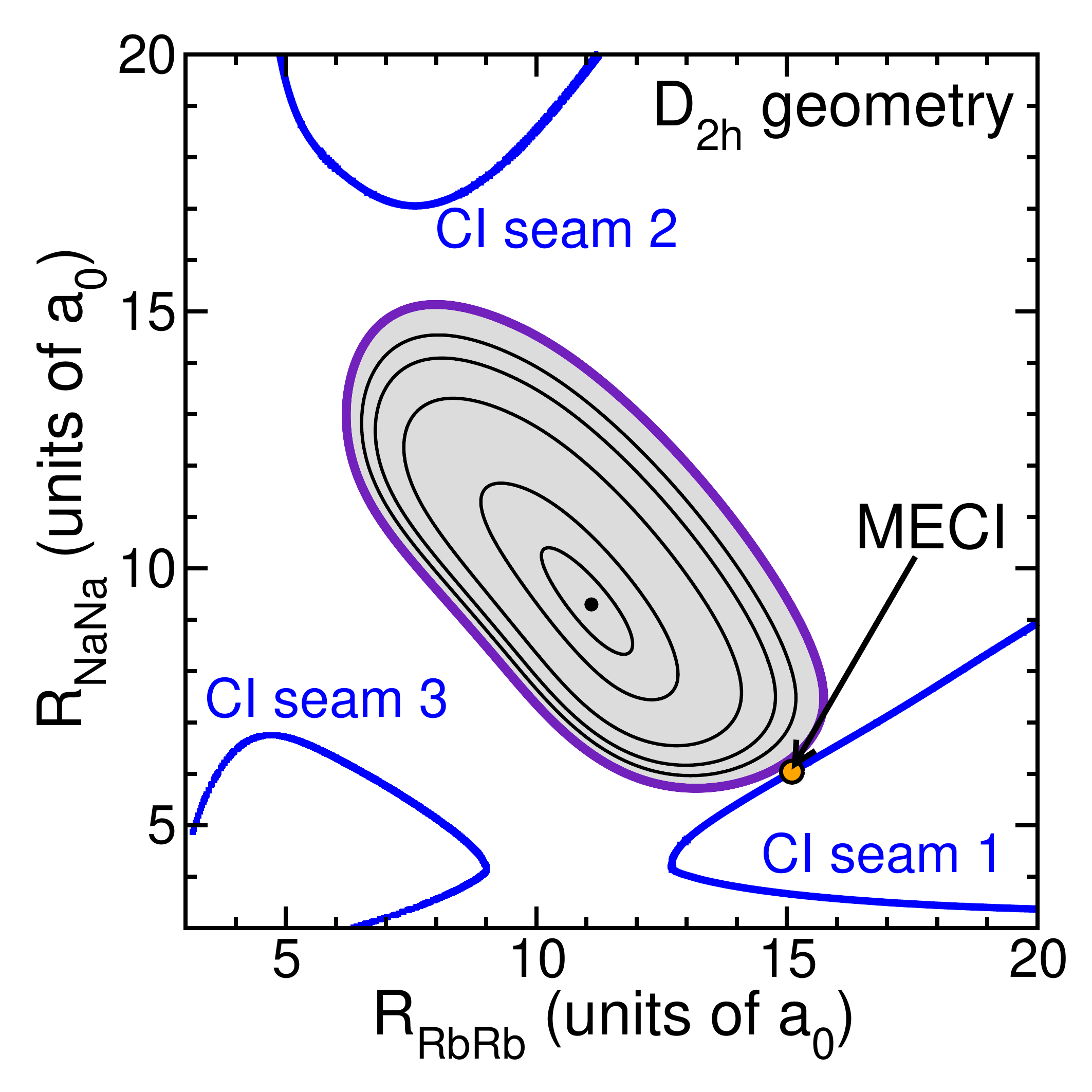} 
      \caption{Contour plot of seams of  conical intersections (thick blue curves) and potential energies (black curves) of  Na$_2$Rb$_2$  in planar $D_{ 2h}$ symmetry as functions of separations $R_{\rm NaNa}$ and $R_{\rm RbRb}$. The orange marker indicates the minimum-energy conical intersection (MECI)  on seam 1. The grey area indicates $D_{2h}$ geometries that are classically accessible for a cold NaRb+NaRb collision. Figure was prepared using Grace-5.1.22 software.}
	\label{fig:seam}
\end{figure}

Figure~\ref{fig:seam} shows that there exist three continuous CI seams in $D_{2h}$ geometries, uniquely
described by the homonuclear pair separations $R_{\rm NaNa}$ and $R_{\rm RbRb}$. 
Two of these seams occur when either the $R_{\rm NaNa}$ or $R_{\rm RbRb}$ is significantly larger than their
dimer ground-state equilibrium separation.
The  CI geometry where the potential energy is lowest on any of the three seams is also shown in the figure.
This minimum-energy conical intersection (MECI) is about $hc\times 500$ cm$^{-1}$ higher in energy than our entrance-channel 
energy and, thus, does not overlap the classically allowed region for the quasi-classical trajectories.
We must conclude that roaming does not require the presence of CIs. The complex landscape of the six-dimensional potential energy surface with its chaotic motion is sufficient for the appearance of roaming.

\vspace*{0.5 cm}
\noindent
{\bf Light-induced transition to excited states.} In this subsection, we describe the time-dependent quasiclassical dressed-state scattering of two cold $^{23}$Na$^{87}$Rb molecules in the presence of trapping laser light with a wavelength of 1064 nm. We also estimate the 
probability  that the laser excites the four-atom complex during the 50 ns mean collisional lifetime found in the previous subsections. 

\begin{figure*}
\includegraphics[scale=0.36]{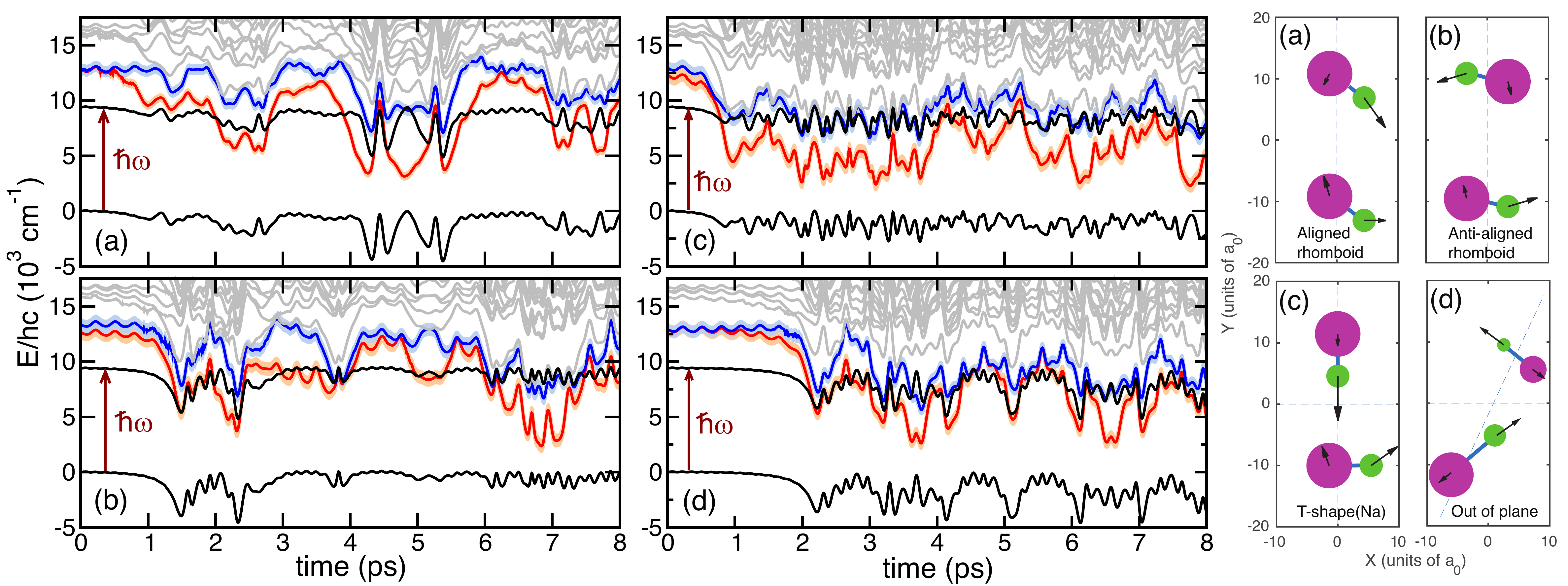}
\caption{{\it On-the-fly} dressed-state model of optical excitations of two classically-colliding cold ground-state 
$^{23}$Na$^{87}$Rb molecules in the presence of an infra-red trapping laser with a wavelength of 1064 nm. 
Panels (a)-(d) show singlet ground- (solid black curve) and excited-states (solid red, blue, and grey curves) potential 
energies as functions of time during the first few ps of trajectories
with zero impact parameter, but different initial relative orientations and velocities of the cold molecules. These initial 
orientations and velocities are shown in the four small panels on the right with corresponding panel labels. Large 
magenta and smaller green filled circles give the positions of Rb and Na atoms, respectively, while arrows of different 
lengths specify their velocities.
The second higher-in-energy black curves in the panels on the left correspond to the ground-state potential plus the energy of a  photon from 
the 1064 nm laser. Excitations occur when this dressed ground-state potential crosses that of excited states.
The error bands on the potentials of the first (orange curve) and second (blue curve) excited states indicate the difference in energy splittings between excited states and the ground state as found by TD-DFT  and  EOM-CCSD calculations. Figures were prepared using Grace-5.1.22 software.}
\label{fig:cycling}
\end{figure*}

For these goals we do  not only need the electronic ground state but also  excited states that have significant electric-dipole 
coupling with the ground state, {\it i.e.} those that are optically-active and can be resonantly excited. The dimer-in-molecule model 
cannot be used to determine all relevant optically-active excited-state potentials and electric dipole moments. Instead, we use 
density functional theory (DFT) to obtain the {\it on-the-fly}  ground-state potential, 
as the four atoms classically evolve under the  forces due to this potential,
and time-dependent density functional theory (TD-DFT) to compute splittings between ground- and excited-state potentials  as well as transition dipole moments. A  description of these electronic-structure calculations is given in Methods. 

Figure \ref{fig:cycling} shows ground- and spin-singlet  excited-state potential energies as functions of time up to 
8 ps for four of our trajectories as well as their initial state sampled from the $v = 0, J = 0$ X$^1\Sigma^+$ state of NaRb.
Their initial relative  kinetic energy  is 1~mK at an initial dimer separation of $38a_0$. The trajectories are 
calculated  using a variable time step size no larger than 4 fs. 
For the trajectories the collision did not finish before we stopped the simulations.
In the figure we also show the energy of the dressed ground state corresponding to the ground-state potential
plus the energy of a 1064 nm photon.  The four-atom complex can be optically-excited when the potential energy of  an
excited state  equals that of the dressed ground state.  In fact, for the 1064 nm laser the potential of only the two 
energetically lowest excited states ``cross'' that of the dressed ground state. The figure shows multiple crossings within our ps 
time windows. 

The computation of {\it on-the-fly} DFT ground and excited potentials is numerically demanding and we had to decrease the 
initial dimer separation from that used in the previous sections and limit the time evolution   to about 10 ps.
Error bands around excited-state potential energies in Fig.~\ref{fig:cycling} are our estimates of the standard-deviation 
uncertainty of splittings between potentials  based on additional EOM-CCSD calculations for the three energetically lowest  
potential surfaces.  

The crossings shown in Fig.~\ref{fig:cycling} immediately suggest a means to determine  the probability of non-adiabatic transitions 
between the potentials. The physics reduces to a time-dependent
two-state system,  considered by Landau  \cite{Landau1932} and Zener \cite{Zener1932}, with diabatic energies that vary 
linearly in time and a  time-independent coupling due to the molecule-laser interaction.
In the spatial domain this corresponds to a transition from one potential to another that is ``vertical'', {\it i.e.} where the atomic positions and velocities are unchanged in the transition.  
Landau-Zener theory then states that the transition probability of excitation from the ground state at the $i^{\rm th}$ crossing, 
occurring at time $t_i$ and $i=1,2,3,\cdots$, is 
 \begin{eqnarray}
 \label{hopping_i}
 p_{i} = \exp\left(-2\pi\frac{\Omega(t_i)^2}{\hbar|\alpha(t_i)|}\right)\,,
 \end{eqnarray}
 where   energy
 \begin{eqnarray}
 \Omega(t_i) = d(t_i) \sqrt{\frac{2I}{\epsilon_0 c}}
\end{eqnarray}
is the coupling matrix element that depends on transition electric  dipole moment $d(t_i)$ to
the relevant excited state and laser intensity $I$.  Here,  $\epsilon_0$ is the vacuum
permittivity.  The rate of change of the energies at  time $t_i$ is $\alpha(t_i)=v_i \beta_i$,
where  $v_i$ is the velocity along the classical path ${\bf R}_{\rm cl}(t)$ of the atomic coordinates at $t=t_i$ and
\begin{equation}
 \beta_i =  {\bf n}_{\rm cl}\cdot \nabla \left(V_{\text{gr}}({\bf R}_{\rm cl}) - V_{\text{exc}}({\bf R}_{\rm cl})\right)
 %\left. \frac{{\rm d} \left(V_{\text{gr}}({\bf R}_{\rm cl}) - V_{\text{exc}}({\bf R}_{\rm cl})\right)}{{\rm d}R_{\rm cl}}\right|_{t=t_i}
\end{equation}
is the spatial derivative of the potential energy difference between the ground- and excited-state potential, $V_{\text{gr}}({\bf R})$ and $V_{\text{exc}}({\bf R})$, respectively, along the classical path at $t=t_i$. Here,  direction ${\bf n}_{\rm cl}={\bf R}_{\rm cl}/|{\bf R}_{\rm cl}|$.

Once excited the atoms  accelerate and move according to the forces generated by the
electronically excited state potential. The excited state can decay back to the ground state
by spontaneous emission or undergo other Landau-Zener transitions, either
back to the ground state or up to doubly-excited states whenever  dressed-state potentials are
resonant. In either case the system's molecular total energy has changed sufficiently from
that of the initial collision state so that the molecules are no longer trapped by the laser
and lost.

For our system typical values of $|\beta(t_i)|$ are $\sim hc\times 10^{-2}\ {\rm
cm}^{-1}/a_0$, while those of $|d(t_i)|$ range from $10^{-3}\,ea_0$ to  $1\,ea_0$, where $e$
is the elementary charge. Moreover, for a laser wavelength of 1064 nm the  mean time interval
between crossings $\delta t=\langle t_{i+1}-t_i\rangle$ is 0.23 ps based on averaging within a
trajectory and our sample of trajectories. In fact, we find that the intervals $t_{i+1}-t_i$
have a Poisson distribution. For a laser at this wavelength and an intensity of 10 kW/cm$^2$,
the mean excitation probability for a crossing $\bar p=\langle p_i\rangle$ is $2.1\times
10^{-5}$, orders of magnitude smaller than one. From Eq.~(\ref{hopping_i}) it follows that
$\bar p$ is proportional to laser intensity. The probability distributions for $t_{i+1}-t_i$
and $p_i$, which we find to be uncorrelated, are not expected to change for propagation times
significantly larger than 10 ps.

As we are unable to run the {\it on-the-fly} calculations out to collision times of order of tens of nanoseconds,
we must derive a coarse grained model for the excitation process based on the inequality $\delta t \ll \tau$. First, we realize that
after $N$ crossings for each trajectory  the total or cumulative likelihood to be excited, and thus lost, is
\begin{equation}
     \sum_{i=1}^N \left[\prod_{j=1}^{i-1}(1-p_j)\right] p_i \,.
\end{equation}
This corresponds to time $t=t_N-t_1=\sum_{i=1}^{N-1}(t_{i+1}-t_i)$ after the first excitation. Next, we assume that $t\gg \delta t$ (and thus $N\gg1$) and sample
$p_i$ and $t_{i+1}-t_i$ from their respective probability distributions. The averaged cumulative excitation probability at time $t$ 
is then well approximated by
\begin{eqnarray}
     P(t) &\approx& 1 - e^{-\bar p t/\delta t}  \equiv 1 - e^{- \kappa t}  
\end{eqnarray}
by replacing $p_i$ and $t_{i+1}-t_i$ by their mean values, using the geometric series to
evaluate the finite sum, and using that $\bar p\ll 1$. On the right hand side of the equation,
we  defined the physically relevant laser-induced survival rate $\kappa=\bar p/\delta t$ of
the collisional complex, which for typical laser intensities is proportional to the laser
intensity.

For the laser with a wavelength of 1064 nm and an intensity of 10 kW/cm$^2$ the  decay time 
$1/\kappa = 12$~ns,  much larger than $\delta t$,  justifying our derivation.  On the other hand the 12-ns decay 
time is on the order of the mean lifetime $\langle \tau\rangle$ of the collisional complex.  
These two characteristic times  of the collisional complex
can be combined to give the likelihood that a NaRb molecule does not survive the collision. This likelihood
is $P(\langle \tau\rangle)$, the averaged cumulative excitation probability at time $\langle \tau\rangle$.   For a 
1064 nm laser at  ${I=10}$ kW/cm$^2$ its value is ${1-0.016}$, close to one. A NaRb molecule is not
likely to survive a collision in the presence of  1064 nm light.

Finally, we  make some observations regarding the relevance of the impact parameter on our conclusions so far.
All reported data  have been for zero impact parameter. From additional trajectory simulations (not shown) we found
that for impact parameters as large as $40a_0$, approximately five times the di-atomic X-state equilibrium separations, the duration 
of the collision and mean excitation probabilities fluctuate by no more than 20\,\%.  (Impact parameters larger than $40a_0$
are  inconsistent with our initial molecular separation and have not been considered.) The independence of the relevant 
quantities with respect to impact parameter is consistent with the theory behind spiraling collisions, Langevin or orbiting 
cross sections, as  well as the chaotic motion in the six-dimensional ground-state potential. Firstly, because 
for low collision energies and the attractive long-range van-der-Waals interaction between two NaRb molecules, the 
maximum expected contributing impact parameter \cite{Child} is several times larger than $40a_0$. Moreover, chaotic
motion with its inherent sampling of all allowed phase space implies that once the short-ranged four-body complex
is formed at the beginning of the collision the typical or mean collision time is independent of impact parameter. Thus,
we can conclude that the light-induced inelastic cross section is approximately $P(\langle \tau\rangle)$ times the Langevin 
cross section.

\vspace*{0.5 cm}
\noindent
{\bf  \large{Discussion}}\\
We have performed a detailed examination of classical collision trajectories in 
$^{23}$Na$^{87}$Rb+$^{23}$Na$^{87}$Rb interactions both with and without the presence of laser light trapping 
the cold molecules. We  computed the mean collision duration of the four-body complex without laser excitations and 
showed evidence of a long‐lived ``collision complex'' that includes roaming motion to briefly form homonuclear 
$^{23}$Na$_2$+$^{87}$Rb$_2$. We also determined the survival or collision time of the four-body complex in the presence of 
1064-nm laser light, which can induce transitions to two electronically excited states, and found that it is of the same order of magnitude as the duration of the four-body complex for a typical experimental laser intensity of 10 kW/cm$^2$. Under these 
circumstances $^{23}$Na$^{87}$Rb molecules will be rapidly removed from the optical trap. 

Our investigation, which includes roaming pathways within the collisional complex, leads to different, much-shorter lifetimes than those from the RRKM theory, which does not have a concept of roaming pathways. We believe that the existence of roaming events during collisions can significantly modify the collisional lifetime and change the reaction outcome. One example of such effect is given in Ref.\cite{Matsugi2013}, which reported the several orders of magnitude decrease of the rate constant for H-atom formation in photodissociation of C$_2$H$_5$ molecule due to the roaming dissociation channel.

Our estimate of the losses induced by the trapping laser assumes that the laser has a nearly uniform spatial intensity profile over size of the molecular cloud as opposed to the experiments of \cite{Bause2021, Gersema2021}, where a chopped ODT with a time duration of laser pulses of 50-100 $\mu$s is used in attempts to decrease the molecular loss rate. Our calculations show that even much shorter collision time of 50 ns is sufficient for colliding molecules to be excited by the trapping light and lost from the trap.  

Several open questions remain regarding the validity of the approximations used in the paper and must be answered
with future investigations. These concerns include the need for full quantum simulations of the collision on the 
ground-state potential, treatment of the CIs, and more sophisticated models for laser excitations. We also envision a 
study of a wavelength dependence of the optical excitation. Longer wavelengths should lead to fewer excitations as, for 
example, excitations to the second excited electronic state become energetically forbidden. Those to the first excited state 
with its CIs with the ground state potential are always possible.
Nevertheless, we feel that the general conclusions will not be invalidated.

\vspace*{0.5 cm}
\noindent
{\bf  \large{Methods}}\\
{\bf Dimer-in-molecule potentials.} For the determination of the mean duration of $^{23}$Na$^{87}$Rb+$^{23}$Na$^{87}$Rb collisions and the tetramer conical intersections we require an easy to compute but at the same time reasonably accurate electron-spin-singlet ground-state PES. We use the  six-dimensional non-relativistic potential based on the dimer-in-molecule theory of Ref.~\cite{Ellison1963}. In this theory the electron wavefunctions of the tetramer are approximated by superpositions of products of four spin-1/2  ground-state alkali-metal-atom electron wavefunctions, such that the total electron spin is zero. Two (geometry- or position-independent) singlet electron wavefunctions can be constructed with straight-forward application of angular momentum algebra \cite{BrinkSatchler}. The corresponding $2\times 2$  tetramer potential matrix is determined from matrix elements for the sum of six pairwise potential operators. Each pairwise potential operator is given by  $\sum_{S_{ij}=0}^1 P_{S_{ij}} V_{S_{ij}}(R_{ij})$,
where $S_{ij}$ is the  electron spin of atom pair $(i,j)$, operator $P_{S_{ij}}$ is the projection operator on states with
pair spin $S_{ij}$ with $P_0+P_1$ equal to the identity operator,
and $V_{S_{ij}}(R_{ij})$ is  the isotropic pair-wise X$^1\Sigma^+$ (or X$^1\Sigma^+_g$) Born-Oppenheimer potential  at
pair separation $R_{ij}$  for $S_{ij}=0$ and the triplet a$^3\Sigma^+$ (or a$^3\Sigma^+_u$) potential for  $S_{ij}=1$.
The three matrix elements of the $2\times2$ matrix are sums of $V_{S_{ij}}(R_{ij})$ with weights given by matrix elements
of a unitary transformation reflecting the different ways to couple  four spin-1/2 atoms into  pairs. These
weights are  proportional to a nine-$j$ symbol and phase factors \cite{BrinkSatchler}.

We use the spectroscopically-accurate dimer potentials from Refs.~\cite{Knoop2011, Strauss2010, Wang2013} with 
adjustments to their repulsive walls at small pair separations, where the potential energy is larger than
that at the dissociation limit.  These adjustments removed unphysical trends going from the light Na$_2$,
to NaRb, to the heavier Rb$_2$ dimers. The extrapolation of spectroscopic data to small separations can not reliably specify 
the shape of the inner walls of potentials.

The energetically-lowest eigenvalue of the $2\times 2$   potential matrix corresponds to the ground-state
tetramer potential and is used in classical trajectory calculations. 
This PES is not separable but, nevertheless, does not include so-called non-additive contributions. We assume that the effects of 
such contributions on the mean lifetime  are small, relying on the results of time-independent quantum-mechanical studies of reactive atom-dimer   collisions with alkali-metal atoms and molecules \cite{Croft2017, Kendrick2020}. They  found that  total reaction rate coefficients are not affected when the non-additive part of the potential is or is not included. 
Similarly, we assume that the lifetimes are   not affected by the precise nature of the adjustments to the  inner walls of the pair
potentials.

Conical intersections in $2\times2$  dimer-in-molecule model  occur when the two eigenvalues of the 
matrix are equal. A necessary condition for such occurrence is that the off-diagonal matrix element
is zero. Inspection of the corresponding weighted sum of $V_{S_{ij}}(R_{ij})$ for Na$_2$Rb$_2$ shows that this matrix element 
is zero for all geometries with $D_{2h}$ symmetry. Thus, for such geometries conical intersections occur
when the diagonal elements of the $2\times2$ matrix are equal. Figure~\ref{fig:seam} shows the locations of the  conical intersections in the dimer-in-molecule model.

\vspace*{0.5 cm}
\noindent
{\bf The {\it on-the-fly}  optical excitation model.} The {\it on-the-fly}  optical excitation model for $^{23}$Na$^{87}$Rb+$^{23}$Na$^{87}$Rb collisions in the presence of a laser field has two steps. First, we  performed classical trajectory 
calculations using the {\it on-the-fly}  ground-state electronic potential energies and gradients from a  density functional theory 
(DFT) method. 
Second, at each of the geometries along a trajectory we  applied time-dependent density functional theory (TD-DFT) calculations to obtain energy splittings  between ground and electronically excited states as well as the corresponding electric dipole moment.
The  molecular dynamics equations  have only been propagated to about 10 ps.

The DFT calculations have been performed for  total electron spin ${S=0}$ using the hybrid functional wb97xd~\cite{Chai2008}. 
This functional is a long-range corrected functional that allows for a good description of the long-range van-der-Waals interaction 
between atoms and molecules. We use a correlation-consistent triple zeta basis  (cc-pvtz) for the Na atom and the 
Stuttgart/Cologne-group effective-core potential ECP28MDF~\cite{Lim2005} for the Rb atom with Hill and Peterson's  augmented 
correlation-consistent quadruple-zeta polarization potential basis (aug-cc-pvqz-pp) \cite{Hill2017}. This choice of basis allows for 
reasonable computational times of  the {\it on-the-fly} dynamics. 
 
The TD-DFT calculations, which found about ten excited states, are based on the  spin-unrestricted Coulomb-attenuated B3LYP functional (uCAM-B3LYP).  We compare the lowest two TD-DFT excitation energies with those from  coupled-cluster equation-of-motion with singles and doubles (EOM-CCSD) calculations using the same basis set and find a nearly 5 \% difference as shown in Fig.~\ref{fig:cycling}. This small difference is sufficient for our purposes.

\vspace*{0.5 cm}
\noindent
{\bf  \large{Acknowledgments}}\\
The authors thank Dr. Maikel Y. Ballester for sharing  the Mercury/Venus-96 code and  helpful suggestions
and Dr. Dajun Wang for fruitful discussions.
S.K., J. K., H. L., and Q. G. acknowledge support from the Army Research Office Grant No. W911NF-17-1-0563, the U.S. Air Force Office of Scientific Research Grant No. FA9550-19-1-0272, and National Science Foundation Grant No. PHY-1908634.

\vspace*{0.5 cm}
\noindent
{\bf  \large{Author contributions}}\\
S.K. conceived, designed and coordinated the work. E.T. and S.K. wrote the manuscript with help of J.K. E.T., J.K., Q.G., M.L., and H.L. have contributed to the development of the theory model and calculations. All authors discussed the results and contributed to the data analysis. Figure were prepared by S.K, E. T., J. K., and H. L.

\bibliographystyle{unsrt}
\bibliography{refs_NaRb-3_with_doi}

\begin{thebibliography}{10}

\bibitem{Ni2008}
K.-K. Ni, S.~Ospelkaus, M.~H.~G. de~Miranda, A.~Pe'er, B.~Neyenhuis, J.~J.
  Zirbel, S.~Kotochigova, P.~S. Julienne, D.~S. Jin, and J.~Ye.
\newblock A high phase-space-density gas of polar molecules.
\newblock {\em Science}, 322:231--235, 2008.

\bibitem{Liu2020}
Y.~Liu, M.-G. Hu, M.~A. Nichols, D.~D. Grimes, T.~Karman, H.~Guo, and K.-K. Ni.
\newblock Photo-excitation of long-lived transient intermediates in ultracold
  reactions.
\newblock {\em Nature Phys.}, 16:1132--1136, 2020.

\bibitem{Takekoshi2014}
T.~Takekoshi, L.~Reichs\"ollner, A.~Schindewolf, J.~M. Hutson, C.~R. Le~Sueur,
  O.~Dulieu, F.~Ferlaino, R.~Grimm, and H.-C. N\"agerl.
\newblock Ultracold dense samples of dipolar $\mathrm{RbCs}$ molecules in the
  rovibrational and hyperfine ground state.
\newblock {\em Phys. Rev. Lett.}, 113:205301, 2014.

\bibitem{Cornish2014}
P.~K. Molony, P.~D. Gregory, Z.~Ji, B.~Lu, M.~P. K\"oppinger, C.~R. Le~Sueur,
  C.~L. Blackley, J.~M. Hutson, and S.~L. Cornish.
\newblock Creation of ultracold $^{87}\mathrm{Rb}^{133}\mathrm{Cs}$ molecules
  in the rovibrational ground state.
\newblock {\em Phys. Rev. Lett.}, 113:255301, 2014.

\bibitem{Park2015}
J.~W. Park, S.~A. Will, and M.~W. Zwierlein.
\newblock Ultracold dipolar gas of fermionic $^{23}\mathrm{Na}^{40}\mathrm{K}$
  molecules in their absolute ground state.
\newblock {\em Phys. Rev. Lett.}, 114:205302, 2015.

\bibitem{Guo2016}
M.~Y. Guo, B.~Zhu, B.~Lu, X.~Ye, F.~D. Wang, R.~Vexiau, N.~Bouloufa-Maafa,
  G.~Qu\'em\'ener, O.~Dulieu, and D.~J. Wang.
\newblock Creation of an ultracold gas of ground-state dipolar
  $^{23}\mathrm{Na}^{87}\mathrm{Rb}$ molecules.
\newblock {\em Phys. Rev. Lett.}, 116:205303, 2016.

\bibitem{Frauke2018}
F.~See\ss{}elberg, N.~Buchheim, Z.-K. Lu, T.~Schneider, X.-Y. Luo, E.~Tiemann,
  I.~Bloch, and C.~Gohle.
\newblock Modeling the adiabatic creation of ultracold polar
  $^{23}\mathrm{Na}^{40}\mathrm{K}$ molecules.
\newblock {\em Phys. Rev. A}, 97:013405, 2018.

\bibitem{Ospelkaus2020}
Kai~K. Voges, Philipp Gersema, Mara Meyer~zum Alten~Borgloh, Torben~A. Schulze,
  Torsten Hartmann, Alessandro Zenesini, and Silke Ospelkaus.
\newblock Ultracold gas of bosonic $^{23}\mathrm{Na}^{39}\mathrm{K}$
  ground-state molecules.
\newblock {\em Phys. Rev. Lett.}, 125:083401, Aug 2020.

\bibitem{Chen2011}
Kuang Chen, Steven~J. Schowalter, Svetlana Kotochigova, Alexander Petrov,
  Wade~G. Rellergert, Scott~T. Sullivan, and Eric~R. Hudson.
\newblock Molecular-ion trap-depletion spectroscopy of {B}a{Cl}${}^{+}$.
\newblock {\em Phys. Rev. A}, 83:030501, Mar 2011.

\bibitem{Sullivan2012}
Scott~T. Sullivan, Wade~G. Rellergert, Svetlana Kotochigova, and Eric~R.
  Hudson.
\newblock Role of electronic excitations in ground-state-forbidden inelastic
  collisions between ultracold atoms and ions.
\newblock {\em Phys. Rev. Lett.}, 109:223002, Nov 2012.

\bibitem{Rellergert2013}
Wade~G. Rellergert, Scott~T. Sullivan, Steven~J. Schowalter, Svetlana
  Kotochigova, Kuang Chen, and Eric~R. Hudson.
\newblock Evidence for sympathetic vibrational cooling of translationally cold
  molecules.
\newblock {\em Nature}, 495:490=494, 2013.

\bibitem{Harter2013}
A.~H\"{a}rter, A.~Kr\"{u}kow, M.~Dei$\beta$, B.~Drews, E.~Tiemann, and
  J.~Hecker Denschlag.
\newblock Population distribution of product states following three-body
  recombination in an ultracold atomic gas.
\newblock {\em Nature Physics}, 9:512--517, 2013.

\bibitem{Mohammadi2021}
Amir Mohammadi, Artjom Kr\"ukow, Amir Mahdian, Markus Dei\ss{}, Jes\'us
  P\'erez-R\'{\i}os, Humberto da~Silva, Maurice Raoult, Olivier Dulieu, and
  Johannes Hecker~Denschlag.
\newblock Life and death of a cold {B}a{R}b$^{+}$ molecule inside an ultracold
  cloud of {R}b atoms.
\newblock {\em Phys. Rev. Research}, 3:013196, Feb 2021.

\bibitem{Krems2008}
R.~V. Krems.
\newblock Cold controlled chemistry.
\newblock {\em Phys. Chem. Chem. Phys.}, 10:4079--4092, 2008.

\bibitem{Martin2009}
Martin~T. B. and Timothy~P. S.
\newblock Ultracold molecules and ultracold chemistry.
\newblock {\em Mol. Phys.}, 107:99--132, 2009.

\bibitem{Dulieu2011}
O.~Dulieu, R.~Krems, M.~Weidemüller, and S.~Willitsch.
\newblock Physics and chemistry of cold molecules.
\newblock {\em Phys. Chem. Chem. Phys.}, 13:18703--18704, 2011.

\bibitem{Balakrishnan2016}
N.~Balakrishnan.
\newblock Perspective: {U}ltracold molecules and the dawn of cold controlled
  chemistry.
\newblock {\em J. Chem. Phys.}, 145:150901, 2016.

\bibitem{Hu2019}
M.-G. Hu, Y.~Liu, D.~D. Grimes, Y.-W. Lin, A.~H. Gheorghe, R.~Vexiau,
  N.~Bouloufa-Maafa, O.~Dulieu, T.~Rosenband, and K.-K. Ni.
\newblock Direct observation of bimolecular reactions of ultracold
  $\mathrm{KRb}$ molecules.
\newblock {\em Science}, 366:1111--1115, 2019.

\bibitem{Hu2020}
M.-G. Hu, Y.~Liu, M.~A. Nichols, L.~B. Zhu, G.~Qu{\'e}m{\'e}ner, O.~Dulieu, and
  K.-K. Ni.
\newblock Product-state control of ultracold reactions via conserved nuclear
  spins.
\newblock {\em arXiv preprint arXiv:2005.10820}, 2020.

\bibitem{Micheli2006}
A.~Micheli, G.~K. Brennen, and P.~Zoller.
\newblock A toolbox for lattice-spin models with polar molecules.
\newblock {\em Nature Physics}, 2:341--347, 2006.

\bibitem{Yan2013}
B.~Yan, S.~A. Moses, B.~Gadway, J.~P. Covey, K.~R.~A. Hazzard, A.~M. Rey, D.~S.
  Jin, and J.~Ye.
\newblock Observation of dipolar spin-exchange interactions with
  lattice-confined polar molecules.
\newblock {\em Nature}, 501:521--525, 2013.

\bibitem{DeMille2002}
D.~DeMille.
\newblock Quantum computation with trapped polar molecules.
\newblock {\em Phys. Rev. Lett.}, 88:067901, 2002.

\bibitem{Ni2018}
K.-K. Ni, T.~Rosenband, and D.~D. Grimes.
\newblock Dipolar exchange quantum logic gate with polar molecules.
\newblock {\em Chem. Sci.}, 9:6830--6838, 2018.

\bibitem{Sawant2020}
R.~Sawant, J.~A. Blackmore, P.~D. Gregory, J.~Mur-Petit, D.~Jaksch,
  J.~Aldegunde, J.~M. Hutson, M.~R. Tarbutt, and S.~L. Cornish.
\newblock Ultracold polar molecules as qudits.
\newblock {\em New J. Phys.}, 22:013027, 2020.

\bibitem{Hughes2020}
M.~Hughes, M.~D. Frye, R.~Sawant, G.~Bhole, J.~A. Jones, S.~L. Cornish, M.~R.
  Tarbutt, J.~M. Hutson, D.~Jaksch, and J.~Mur-Petit.
\newblock Robust entangling gate for polar molecules using magnetic and
  microwave fields.
\newblock {\em Phys. Rev. A}, 101:062308, 2020.

\bibitem{Zuchowski2010}
P.~S. \ifmmode~\dot{Z}\else \.{Z}\fi{}uchowski and J.~M. Hutson.
\newblock Reactions of ultracold alkali-metal dimers.
\newblock {\em Phys. Rev. A}, 81:060703, 2010.

\bibitem{Ospelkaus2010a}
S.~Ospelkaus, K.-K. Ni, D.~Wang, M.~H.~G. de~Miranda, B.~Neyenhuis,
  G.~Qu{\'e}m{\'e}ner, P.~S. Julienne, J.~L. Bohn, D.~S. Jin, and J.~Ye.
\newblock Quantum-state controlled chemical reactions of ultracold
  potassium-rubidium molecules.
\newblock {\em Science}, 327:853--857, 2010.

\bibitem{Idziaszek2010}
Z.~Idziaszek and P.~S. Julienne.
\newblock Universal rate constants for reactive collisions of ultracold
  molecules.
\newblock {\em Phys. Rev. Lett.}, 104:113202, 2010.

\bibitem{Kotochigova2010_1}
S.~Kotochigova.
\newblock Dispersion interactions and reactive collisions of ultracold polar
  molecules.
\newblock {\em New J. Phys.}, 12:073041, 2010.

\bibitem{DeMarco2019}
L.~De~Marco, G.~Valtolina, K.~Matsuda, W.~G. Tobias, J.~P. Covey, and J.~Ye.
\newblock A degenerate {F}ermi gas of polar molecules.
\newblock {\em Science}, 363:853--856, 2019.

\bibitem{Ye2018}
X.~Ye, M.~Guo, M.~L. Gonz{\'a}lez-Mart{\'\i}nez, G.~Qu{\'e}m{\'e}ner, and
  D.~Wang.
\newblock Collisions of ultracold $^{23}\mathrm{Na}^{87}\mathrm{Rb}$ molecules
  with controlled chemical reactivities.
\newblock {\em Sci. Adv.}, 4, 2018.

\bibitem{Gregory2019}
P.~D. Gregory, M.~D. Frye, J.~A. Blackmore, E.~M. Bridge, R.~Sawant, J.~M.
  Hutson, and S.~L. Cornish.
\newblock Sticky collisions of ultracold $\mathrm{RbCs}$ molecules.
\newblock {\em Nat. Commun.}, 10:1--7, 2019.

\bibitem{Bause2021}
Roman~Bause et~al.
\newblock Collisions of ultracold molecules in bright and dark optical dipole
  traps.
\newblock {\em arXiv preprint arXiv:2103.00889v2}, 2021.

\bibitem{Gersema2021}
Philipp~Gersema et~al.
\newblock Probing photoinduced two-body loss of ultracold non-reactive bosonic
  $^{23}${N}a$^{87}${R}b and $^{23}${N}a$^{39}${K}molecules.
\newblock {\em arXiv preprint arXiv:2103.00510v2}, 2021.

\bibitem{Christianen2019}
A.~Christianen, M.~W. Zwierlein, G.~C. Groenenboom, and T.~Karman.
\newblock Photoinduced two-body loss of ultracold molecules.
\newblock {\em Phys. Rev. Lett.}, 123:123402, 2019.

\bibitem{Gregory2020}
P.~D. Gregory, J.~A. Blackmore, S.~L. Bromley, and S.~L. Cornish.
\newblock Loss of ultracold $^{87}\mathrm{Rb}^{133}\mathrm{Cs}$ molecules via
  optical excitation of long-lived two-body collision complexes.
\newblock {\em Phys. Rev. Lett.}, 124:163402, 2020.

\bibitem{Softley2009}
M.~T. Bell and T.~P. Softley.
\newblock Ultracold molecules and ultracold chemistry.
\newblock {\em Mol. Phys.}, 107:99--132, 2009.

\bibitem{Levine2009}
R.~D. Levine.
\newblock {\em Molecular Reaction Dynamics}.
\newblock Cambridge University Press, 2005.

\bibitem{Ellison1963}
F.~Ellison.
\newblock A method of diatomics in molecules. {I}. general theory and
  application to {H}$_2${O}.
\newblock {\em J. Am. Chem. Soc.}, 85:3540–--3544, 1962.

\bibitem{Hase1996}
W.~L. Hase, R.J. Duchovic, X.~Hu, A.~Komornik, K.F. Lim, D.-H. Lu, G.H.
  Peslherbe, K.N. Swamy, S.R. van~de Linde, A.~J.C. Varandas, and R.J. H.~Wang.
\newblock {MERCURY}: {A} general {M}onte {C}arlo classical trajectory computer
  program.
\newblock {\em QCPE Bull.}, 16:43, 1996.

\bibitem{Domcke2011}
W.~Domcke, D.~Yarkony, and H.~K{\"o}ppel.
\newblock {\em Conical Intersections: {T}heory, Computation and Experiment},
  volume~17 of {\em Advanced series in physical chemistry}.
\newblock World Scientific, Singapore, 2011.

\bibitem{Bowman2014}
J.~M. Bowman.
\newblock Roaming.
\newblock {\em Molecular Physics}, 112:2516--2528, 2014.

\bibitem{Townsend2004}
D.~Townsend, S.~A. Lahankar, Lee~S. K., S.~D. Chambreau, A.~G. Suits, and
  et~al.
\newblock The roaming atom: {S}traying from the reaction path in formaldehyde
  decomposition.
\newblock {\em Science}, 306:1158, 2004.

\bibitem{Christoffel2009}
K.M. Christoffel and J.M. Bowman.
\newblock Three reaction pathways in the {H} + {HCO} $\to$ {H}$_2$ + {CO}
  reaction.
\newblock {\em J. Phys. Chem. A}, 113:4138, 2009.

\bibitem{Takayanagi2011}
T.~Takayanagi and T.~Tanaka.
\newblock Roaming dynamics in the {M}g{H} + {H} $\to$ {M}g + {H}$_2$ reaction:
  {Q}uantum dynamics calculations.
\newblock {\em Chem. Phys. Lett.}, 504:130, 2011.

\bibitem{Guo2013}
A.~Li, J.~Li, and H.~Guo.
\newblock Quantum manifestation of roaming in {H}+ {M}g{H} $\to$ {M}g +
  {H}$_2$: The birth of roaming resonances.
\newblock {\em J. Phys. Chem. A}, 117:5052--5060, 2013.

\bibitem{Bowman2020}
M.~S. Quinn, K.~Nauta, M.~J.~T. Jordan, J.M. Bowman, P.~L. Houston, and S.~H.
  Kable.
\newblock Rotating resonances in the {H}$_2${CO} roaming reaction are revealed
  by detailed correlations.
\newblock {\em Science}, 369:1592--1596, 2020.

\bibitem{Suits2008}
A.~G. Suits.
\newblock Roaming atoms and radicals: {A} new mechanism in molecular
  dissociation.
\newblock {\em Acc. Chem. Res.}, 41:873, 2008.

\bibitem{Herath2011}
N.~Herath and A.~G. Suits.
\newblock Roaming radical reactions.
\newblock {\em J. Phys. Chem. Lett.}, 2:642, 2011.

\bibitem{Shepler2011}
J.~M. Bowman and B.~C. Shepler.
\newblock Roaming radicals.
\newblock {\em Annu. Rev. Phys. Chem.}, 62:531, 2011.

\bibitem{Hase2017}
S.~Pratihar, X.~Ma, Z.~Homayoon, G.~L. Barnes, and W.~L. Hase.
\newblock Direct chemical dynamics simulations.
\newblock {\em J. Am. Chem. Soc.}, 139:3570--3590, 2017.

\bibitem{Landau1932}
L.~D. Landau.
\newblock Zur theorie der {E}nergie\"{u}bertragung. {II}.
\newblock {\em Physics of the Soviet Union}, 2:46--51, 1932.

\bibitem{Zener1932}
C.~Zener.
\newblock Non-adiabatic crossing of energy levels.
\newblock {\em R. Soc. London A}, 137:696--702, 1932.

\bibitem{Weigend2005}
F.~Weigend and R.~Ahlrichs.
\newblock {\em Phys. Chem. Chem. Phys.}, 7:3297, 2005.

\bibitem{Cote2012}
J.~N. Byrd, H.~Harvey~Michels, J.~A. Montgomery, R.~Cote, and W.~C. Stwalley.
\newblock Structure, energetics, and reactions of alkali tetramers.
\newblock {\em J. Chem. Phys.}, 136:014306, 2012.

\bibitem{Harding2012}
L.~B. Harding, S.~J. Klippenstein, and A.~W. Jasper.
\newblock Separability of tight and roaming pathways to molecular
  decomposition.
\newblock {\em J. Phys. Chem. A}, 116:6967--6982, 2012.

\bibitem{Garrido2019}
J.~D. Garrido, S.~Ellakkis, and M.~Y. Ballester.
\newblock Relaxation of vibrationally excited {OH} radical by {SO}.
\newblock {\em J. Phys. Chem. A.}, 123(42):8994--9007, 2019.

\bibitem{Bowman2011}
J.~M. Bowman and A.~G. Suits.
\newblock Roaming reactions: The third way.
\newblock {\em Physics Today}, 11:33--37, 2011.

\bibitem{Nakamura2014}
M.~Nakamura, P.-Y. Tsai, T.~Kasai, K.-C. Lin, F.~Palazzetti, A.~Lombardi, and
  V.~Aquilanti.
\newblock Dynamical, spectroscopic and computational imaging of bond breaking
  in photodissociation: {R}oaming and role of conical intersections.
\newblock {\em Faraday Discussions}, 177:77, 2014.

\bibitem{Maeda2015}
S.~Maeda, T.~Taketsugu, K.~Ohno, and K.~Morokuma.
\newblock From roaming atoms to hopping surfaces: {M}apping out global reaction
  routes in photochemistry.
\newblock {\em J. Am. Chem. Soc.}, 137:3433--3445, 2015.

\bibitem{Wallis2009}
A.~O.~G. Wallis, S.~A. Gardiner, and J.~M. Hutson.
\newblock Conical intersections in laboratory coordinates with ultracold
  molecules.
\newblock {\em Phys. Rev. Lett.}, 103:083201, 2009.

\bibitem{Child}
M.~S. Child.
\newblock {\em Molecular collision theory}.
\newblock Academic Press, London, New York, 1974.

\bibitem{Matsugi2013}
Akira Matsugi.
\newblock Roaming dissociation of ethyl radicals.
\newblock {\em J. Phys. Chem. Lett.}, 4:4237--4240, 2013.

\bibitem{BrinkSatchler}
D.~M. Brink and G.~R. Satchler.
\newblock {\em Angular Momentum}.
\newblock Oxford University Press, Oxford, 3$^{\rm rd}$ edition, 1993.

\bibitem{Knoop2011}
S.~Knoop, T.~Schuster, R.~Scelle, A.~Trautmann, J.~Appmeier, M.~K. Oberthaler,
  E.~Tiesinga, and E.~Tiemann.
\newblock Feshbach spectroscopy and analysis of the interaction potentials of
  ultracold sodium.
\newblock {\em Phys. Rev. A}, 83:042704, 2011.

\bibitem{Strauss2010}
C.~Strauss, T.~Takekoshi, F.~Lang, K.~Winkler, R.~Grimm, J.~Hecker~Denschlag,
  and E.~Tiemann.
\newblock Hyperfine, rotational, and vibrational structure of the
  ${a}^{3}{\ensuremath{\Sigma}}_{u}^{+}$ state of $^{87}\mathrm{Rb}$${}_{2}$.
\newblock {\em Phys. Rev. A}, 82:052514, Nov 2010.

\bibitem{Wang2013}
F.~Wang, D.~Xiong, X.~Li, D.~Wang, and E.~Tiemann.
\newblock Observation of feshbach resonances between ultracold {N}a and {R}b
  atoms.
\newblock {\em Phys. Rev. A}, 87:050702, May 2013.

\bibitem{Croft2017}
J.~F.~E. Croft, C.~Makrides, M.~Li, A.~Petrov, B.~K. Kendrick, N.~Balakrishnan,
  and S.~Kotochigova.
\newblock Universality and chaoticity in ultracold {K}+{KR}b chemical
  reactions.
\newblock {\em Nature Comm.}, 8:15897, 2017.

\bibitem{Kendrick2020}
B.~K. Kendrick, H.~Li, M.~Li, S.~Kotochigova, J.~F.~E. Croft, and
  N.~Balakrishnan.
\newblock Non-adiabatic quantum interference in the ultracold {L}i + {L}i{N}a
  $\to$ {L}i$_2$ + {N}a reaction.
\newblock {\em Accepted by Phys. Chem. Chem. Phys.}, 2021.
\newblock See also arxiv.org/abs/2006.15238.

\bibitem{Chai2008}
J.-D. Chai and M.~Head-Gordon.
\newblock Long-range corrected hybrid density functionals with damped
  atom–atom dispersion corrections.
\newblock {\em Phys. Chem. Chem. Phys.}, 10:6615--6620, 2008.

\bibitem{Lim2005}
I.~S. Lim, P.~Schwerdtfeger, B.~Metz, and H.~Stoll.
\newblock All-electron and relativistic pseudopotential studies for the group
  {I} element polarizabilities from $\mathrm{K}$ to element 119.
\newblock {\em J. Chem. Phys.}, 122:104103, 2005.

\bibitem{Hill2017}
J.~G. Hill and K.~A. Peterson.
\newblock Gaussian basis sets for use in correlated molecular calculations.
  {XI}. {P}seudopotential-based and all-electron relativistic basis sets for
  alkali metal ($\mathrm{K–Fr}$) and alkaline earth ($\mathrm{Ca–Ra}$)
  elements.
\newblock {\em J. Chem. Phys.}, 147:244106, 2017.

\end{thebibliography}

\end{document}